\documentstyle[preprint,eqsecnum,aps,prb,amstex]{revtex}
\begin{document}
\title{Non-linear electrical response \\
in a non-charge-ordered manganite : $\Pr_{0.8}$Ca$_{0.2}$MnO$_3$}
\author{S.\ Mercone, A.\ Wahl$^{*}$, Ch. Simon, C.\ Martin}
\address{Laboratoire CRISMAT, Unit\'{e} Mixte de Recherches 6508, Institut des\\
Sciences de la Mati\`{e}re et du Rayonnement - Universit\'{e} de Caen, 6\\
Boulevard du\\
Mar\'{e}chal Juin, 14050 Caen Cedex, France.}
\date{\today}
\maketitle

\begin{abstract}
Up to now, electric field induced non-linear conduction in the $\Pr_{1-x}$Ca$%
_x$MnO$_3$ system has been ascribed to a current-induced destabilization of
the charge ordered phase. However, for $x\leq 0.25$, a ferromagnetic
insulator state is observed and charge-ordering is absent whatever the
temperature. A systematic investigation of the non-linear transport in the
ferromagnetic insulator Pr$_{0.8}$Ca$_{0.2}$MnO$_3$ shows rather similar
results to those obtained in charge ordered systems. However, the
experimental features observed in Pr$_{0.8}$Ca$_{0.2}$MnO$_3$ are distinct
in that the collapse of the CO energy gap can not be invoked as usually done
in the other members of the PCMO system. We propose interpretations in which
the effectiveness of the DE is restored upon application of electric field.
\end{abstract}


Hole-doped perovskite manganese oxides R$_{1-x}$AE$_x$MnO$_3$ (R and AE,
being trivalent rare-earth and divalent ions, respectively) are associated
with a wide variety of electronic and magnetic properties depending on the
value of $x$ and the averaged $A$-site cation radius, $\langle r_A\rangle $.%
\cite{REVIEW} These materials have recently been the subject of intense
studies due to intriguing phenomena such as charge/orbital ordering (CO)\cite
{CO} or colossal magnetoresistance (CMR).\cite{CMR} The latter is usually
interpreted by means of the double-exchange interaction (DE) scenario\cite
{DE} which gives an interesting qualitative interpretation of coupled
ferromagnetic ordering and metallicity. Within such a framework, the
ferromagnetic (FM) ordering is related to a large electronic itinerancy i.e.
metallic behavior. Among the various mixed valent manganites studied so far,
the $\Pr_{1-x}$Ca$_x$MnO$_3$ system (PCMO) is perhaps the most interesting
because it shows a great variety of ordered phases very sensitive to cation
/ anion doping.\cite{SMO00,LEE95,LEE99,TOM96,MAR99,MAI97,JIR85} For $0.3\leq
x<0.8$, charge ordering of $Mn^{3+}$and $Mn^{4+}$ (CO) is found and an
antiferromagnetic (AFM) ordering can be observed with N\'{e}el temperature
ranging from 100K to 170K for $x=0.8$ and $0.3$, respectively. Due to its
low tolerance factor, this system happens to remain insulating giving rise,
under zero field, to a charge ordered insulating state (COI).\ It has been
widely shown that this COI state in the PCMO system can be melted into a
metallic ferromagnet (FMM) upon application of a magnetic field of
sufficient amplitude.\cite{LEE95,TOM96,YOS95,YOS96,TOM95,ANA99}\ Within the
doping range $0.3\leq x\leq 0.5$, in addition to the magnetic field, such a
destabilization of the COI state can also be induced by external
perturbations such as irradiation by X-ray\cite{KIR97} or light\cite
{FIE98,OGA98} and application of pressure\cite{MOR97} or electric field.\cite
{ASA97,STA00,GUH00a,GUH00b,RAO00,BUD01}\ Those optical or irradiation
induced transitions are usually argued to be a result of classical
percolation transport in a non-homogeneous medium ; however, the role of the
application of electric field is not so clear.

For instance, Guha et $al.$\cite{GUH00a,GUH00b} found that the current
induced destabilization of the COI state in $\Pr_{0.63}$Ca$_{0.37}$MnO$_3$
leads to a regime of negative differential resistivity (NDR) ($\frac{dV}{dI}%
<0)$ with a concomitant enhancement of magnetization. The authors invoke the
creation of low-resistivity conducting paths made up of FM phase along the
current flow. Very recently, Yuzhelevski et $al.$\cite{YUZ01} have claimed
in low doped LCMO system that the influence of light, X-ray and current may
be interpreted in terms of spin-polarized tunnel conduction mechanism
modifying phase separation-conditions along the percolation path.\ 

Up to now, all experiments investigating the effect of electric field on the
conducting behavior of the PCMO system have mostly been carried out on CO
candidates. Consequently, interpretations based on multiphase coexistence
(AF-COI and FMM phases whose respective volumes in the bulk are modified
under perturbations) is often raised.\cite
{FIE98,OGA98,ASA97,STA00,GUH00a,GUH00b} The very low doping regime of the
PCMO system has not been explored yet. For $x\leq 0.25$, a ferromagnetic
insulator (FMI) state is found and charge-ordering is not observed whatever
the temperature. Moreover, a metallic state is never realized, even upon
application of a magnetic field. This can be understood as follows : on the
one hand, for this system, we have an important inward tilt of the $%
Mn^{3+}-O-Mn^{4+}$ bounds which results in a decrease of the d$%
_{z^2}-p\sigma -$d$_{z^2}$ overlaps. This produces a decrease of the
magnitude of the $e_g$ bandwidth and, consequently, in a reduction in the
effectiveness of the double exchange. On the other hand, for $x\leq 0.25$ we
deal with a very low hole concentration which does not favor a delocalized
electronic state.

The aim of our experience was to investigate whether a current-induced
metal-insulator transition can be observed in a non-CO manganite of the PCMO
system. For this purpose, a crystal of $\Pr_{0.8}$Ca$_{0.2}$MnO$_3$, not so
far from the CO region, has been chosen. Although a M-I transition has not
been clearly observed, the resistance for increasing bias current is
drastically reduced.\ Moreover, strong non-linear electrical conduction is
found with a NDR developing for $T<T_c$.\ This behavior is not sensitive
upon application of a magnetic field.\ Compared to already reported results,
the experimental features observed here are distinct in that the collapse of
the CO energy gap can not be invoke as usually done in the other members of
the PCMO system.\cite{FIE98,OGA98,ASA97,STA00,GUH00a,GUH00b,RAO00,BUD01}

Using the floating-zone method with a feeding rod of nominal composition Pr$%
_{0.8}$Ca$_{0.2}$MnO$_3$, a several-cm-long single crystal was grown in a
mirror furnace. Two samples were cut out of the central part of this
crystal, one of them for resistivity measurements and the other for
magnetization and specific heat measurements. X-ray diffraction and electron
diffraction studies, which were performed on pieces coming from the same
part of the crystal, attested that the samples are single phased, and well
crystallized. The cell is orthorhombic with a Pnma space group, in agreement
with previously reported structural data. The energy dispersive spectroscopy
analyses confirm that the composition is homogeneous and close to the
nominal one, in the limit of the accuracy of the technique. The electron
diffraction characterization was also carried out versus temperature, from
room temperature to 90K. The reconstruction of the reciprocal space showed
that the cell parameters and symmetry remain unchanged in the whole domain
of temperature and, more especially, no extra reflections have been
detected. This electron diffraction observation, coupled with lattice
imaging, shows that, in our sample there is no charge ordering effect, even
at short range distances. All X-ray and electron diffraction observations
agree with previous published results for compounds of the same system.\cite
{JIR85}

In order to characterize the magnetic structure of the sample, we have
performed a temperature dependence of the diffraction patterns on the $G4.1$
neutron spectrometer $\left( \lambda =0.2426\ nm\right) $ of the Orphee
source (LLB-Saclay, France) in the 2$\theta $ angular range (2.00 - 81.90).
The sample was first powdered and the diffracted spectra were recorded as
function of temperature from $300K$ down to $1.5K$. For $T<T_c$, the data
were fitted with a $Pnma$ structure with a ferromagnetic order with the $Mn$
spins aligned along the a axis and a resolution limited correlation length
(more than $100nm$). The saturated moment is $3.46(4)$ $\mu _B/Mn$ at $1.5K$
and the Curie temperature is around $135K$. No evidence of any
antiferromagnetic peak was observed in this sample down to $1.5K$. The $1.5K$
experimental and calculated patterns given in Figure 1 test the validity of
the fit. The 020 peak intensity versus temperature (see inset) shows the FM
component evolution and allows $T_c$ determination. A magnetic contribution
was also added for $\Pr $ in the calculation at lower temperatures.

Four linear contact pads of In were soldered onto the sample in linear
four-probe configuration.\ $V-I$ data were taken with current biasing
(Keithley 236) and with a temperature control of $100$ $mK$.\ The
measurements under magnetic field were done using a superconducting magnet
capable of producing $9$ $Tesla$.\ 

In Figure 2, we show the temperature variation of the resistance ($R=\frac VI%
)$ of a Pr$_{0.8}$Ca$_{0.2}$MnO$_3$ crystal for various bias current, $%
R_I(T),$ under zero field. When the current is small ($10^{-3}mA$ et $1mA$),
the sample shows an insulating behavior and, at low temperature, the large
resistance increase overloads our current source .\ Our current source
overloads for $100V$, thus the maximum resistance that we can measure for $%
10^{-3}mA$ is around $100M\Omega $.\ For higher currents ($10mA$ et $50mA$),
the resistance is strongly depressed with a trend to saturation when the
temperature is lowered. We do not see a clear metal-like decrease in
resistance below $T_c$ and the observed behavior is rather similar to that
of Gd$_{0.5}$Ca$_{0.5}$MnO$_3$ and Nd$_{0.5}$Ca$_{0.5}$MnO$_3$ films
deposited on LAO\cite{RAO00} and PCMO crystals.\cite{GUH00a,GUH00b} The
latter papers report that electrical current triggers the collapse of the
low temperature electrically insulating CO state to a FMM state.
Measurements of $R_I(T)$ have been carried out upon cooling and warming and
no hysteresis has been observed. Around $50K$, a slight anomaly can be
observed on every $R_I(T)$, this can be linked to the $\Pr $ magnetic
ordering as observed by neutron experiments.

Figure 3 shows $V-I$ characteristics in a semi-log scale for $T<T_c$ $\left(
80K,90K\text{ and }100K\right) $ under zero field. The $V-I$ characteristics
for temperatures above the Curie temperature $\left( 170K\text{ and }%
300K\right) $ are displayed in Figure 4. For $T=300K$, an ohmic conduction
is observed on the whole current range. As one approaches $T_c$, non-linear
conduction sets in. As $T$ is lowered below $T_c$ the non-linearity
increases and the $V-I$ curves exhibit a negative differential resistance
(NDR).\ The region of NDR is observed when the bias current attains a
current threshold $\left( I_{th}\right) $. The latter is higher when the
temperature is close to $T_c$ from below $\left( 0.4mA\text{ at }80K\text{
up to }2.5mA\text{ at }100K\right) .$ The data displayed in Figures 3 and 4
are highly reproducible and do not show any significant hysteretic behavior
when the bias current is cycled ($5\%$ at its maximum).\ 

We have carefully checked that the Joule heating is irrelevant to account
for this current induced effect.\ The temperature rise of the sample with
respect to the sample holder $\left( \Delta T\right) $ has been measured by
attaching a thermometer on the top of the sample itself. We have obtained $%
\Delta T\leq 25K$ at the lowest temperature $\left( 80K\right) $ and for the
highest power dissipation level. In this low temperature range, the power
dissipation level where the NDR sets in leads to $\Delta T\prec 3K$. For
higher temperature of measurements, $\Delta T$ becomes negligible. Moreover,
for these temperatures, the heat dissipation is low which may induce high
Joule heating and possible non-linear $V-I$ characteristics. However, at $%
300K$, $V-I$ curve is linear on the whole current range. Finally, all these
points confirm that the phenomenons observed above are not triggered by
simple thermal effects or by heating of the crystal.

The $V-I$ characteristics are not strongly modified under magnetic field.\
As an example, Figure 5 shows measurements at $90K$ with and without
magnetic field. It can be observed that the NDR region sets in for rather
close values of the current threshold.. The variation is not substantial
enough to invoke a magnetic field effect. To sum up, the main results are :
(i) Occurence of a strong non-linear conduction as one approaches $T_c$
leading to NDR when the temperature is lowered well below $T_c$.\ (ii)
Non-hysteretic $V-I$ characteristics upon cycling the bias current. (iii) No
modification of the above features when a magnetic field is applied.\ 

The zero field data shown for $\Pr_{0.8}$Ca$_{0.2}$MnO$_3$ are reminiscent
of M-I transition and non-linear conduction induced by electric field in the
CO compounds of the PCMO system ($0.3\leq x<0.8)$.\cite
{ASA97,STA00,GUH00a,GUH00b,RAO00,BUD01} In the latter experiments, a melting
of the CO state is usually invoked to account for this phenomenon.\ In the
case of a current-induced transition, it is proposed that a breakdown of the
CO state locally leads to the creation of metallic filaments made up of FM
phase when a current threshold is passing through the sample. The resistance
drop and, concomitantly, the occurrence of NDR is the result of the current
being drawn into the metallic conducting path forming a parallel circuit.\
Within such a scenario, the experimentally observed disappearance of the NDR
upon application of magnetic field is explained by considering the magnetic
field induced transformation of AF-COI state into FMM. Coexistence of CO
phase and FM one would prevent the nucleation of metallic filaments needed
for occurrence of NDR.\cite{ASA97,GUH00a}

Let's now turn to the data obtained for our weakly doped FMI $\Pr_{0.8}$Ca$%
_{0.2}$MnO$_3$ crystal. As developed in Ref.[11] and on the basis of our
neutron diffraction data charge ordering is never observed in $\Pr_{0.8}$Ca$%
_{0.2}$MnO$_3$. Thus, one can hardly invoke the electric field-induced
destabilization of the CO state to understand our experimental data. Our
results suggest that for a sufficient bias current, a new conduction path is
opened. This conduction path is closely linked to the FM ground state of the
sample and magnetic field does not triggers or limit this phenomenon.
Moreover, the non-hysteretic nature of the $V-I$ characteristics is not in
favor of an electric field-induced metastable state.

Although a theoretical understanding of the mechanisms at work in mixed
valent manganites is still incomplete, the DE is commonly adopted as a main
ingredient. In the completely ferromagnetic phase, $e_g$ electrons of $%
Mn^{3+}$ can hop coherently without magnetic scattering by the $t_{2g}$
spins while they become strongly incoherent if the $t_{2g}$ are disordered.
However, for ferromagnetic mixed manganites, localization effects can arise
due to spatial fluctuations of structural and spin dependent potentials as
discussed by Coey et $al.$\cite{COE95}\ As emphasized before, in $\Pr_{0.8}$%
Ca$_{0.2}$MnO$_3$, the important inward tilt of the $Mn^{3+}-O-Mn^{4+}$
bounds results in a decrease of the $e_g$ bandwidth. An electric field may
induce a local electrical moment in the $MnO_6$ octahedra by modifying the
spatial distribution of the charges. This leads to an increase of the d$%
_{z^2}-p\sigma -$d$_{z^2}$ overlaps i.e. an increase of the $e_g$ bandwidth.
As a consequence, the hopping probability and the mobility of the carriers
would be greatly enhance.\ The magnetic ground state of $\Pr_{0.8}$Ca$_{0.2}$%
MnO$_3$ is definitely ferromagnetic, thus there is no need of a magnetic
field to initiate a DE-mediated hopping. The slight modification between the 
$V-I$ curves with and without magnetic field could explained as a field
induced suppression of the deviation from perfect collinear ferromagnetism
which implies a spin-only moment of $3.8$ $\mu _B/Mn$ (Neutron diffraction
experiments give $3.46(4)$ $\mu _B/Mn$ for $H=0$ $Tesla$). Moreover, this
scenario is supported by the temperature dependence of $I_{th}$. Indeed, the
FM correlations are enhanced as the temperature is lowered well below $T_c$
and smaller bias current threshold may be required to induce delocalization
via DE.

The non-linear conduction, and more particularly the NDR, can also arise as
a result of inelastic tunneling. One can start off from this point by
proposing another kind of process which could be at work in such a low doped
system. To do so, one could imagine a rather inhomogeneous material,
magnetically speaking. In this scenario, due to the random distribution of
the $Mn^{3+}$ and $Mn^{4+}$ ions, the long range FM ordering may appear in
spatially distinct regions, strongly topologically disordered, of the
sample. In those regions, the DE would be locally at work however
metallicity could not be macroscopically observed because these
regions,where the carriers are mobile, would be interrupted by tunnel-type
weak links, whose nature remains unclear. Twinning might be a clue ; such
disorder effects are unavoidable in crystals and could play the role of
tunneling junctions between highly ferromagnetic domains. Thus, the peculiar
behavior observed in $\Pr_{0.8}$Ca$_{0.2}$MnO$_3$ can be tentatively
understood by considering a spin polarized current flowing across
twin-boundary tunnel junctions separating neigbhoring ferromagnetic domains.

In summary, we have carried out a systematic investigation of the non-linear
transport in the FMI $\Pr_{0.8}$Ca$_{0.2}$MnO$_3$. Experimental data are
rather similar to those obtained in charge ordered systems. However, the
ground state of $\Pr_{0.8}$Ca$_{0.2}$MnO$_3$ prevents us to invoke a
current-induced destabilization of the charge ordered phase to account for
experimental data. We propose interpretations in which the effectiveness of
the DE is restored upon application of electric field.

\begin{quote}
Acknowledgements : The authors thank L.\ Herv\'{e} for crystal growth, G.\
Andr\'{e} and N.\ Guiblin for preliminary neutron diffraction results and
M.\ Hervieu for electron diffraction characterization.

* : corresponding author
\end{quote}

\section{Figures Captions}

\begin{description}
\item[Figure 1]  : $1.5K$ neutron diffraction patterns : continuous line for
calculated plot, stars for experimental data, the upper Bragg strikes
correspond to the $Pnma$ crystallographic structure and the lower one are
for the magnetic phase. The main indexations are also given. Inset : 020 FM
peak intensity versus temperature.

\item[Figure 2]  : Temperature dependence of resistance for a $\Pr_{0.8}$Ca$%
_{0.2}$MnO$_3$crystal with various bias currents.

\item[Figure 3]  : $V-I$ characteristics under zero field for temperatures
below $T_c$ ($80K,$ $90K$ and $100K).$

\item[Figure 4]  : $V-I$ characteristics under zero field for temperatures
above $T_c$ ($170K$ and $300K).$

\item[Figure 5]  : $V-I$ characteristics for $T=90K$ under 0 and 8 Tesla.\ 
\end{description}

\end{document}